\documentclass{interact}

\usepackage{graphicx}%
\usepackage{multirow}%
\usepackage{amsmath,amssymb,amsfonts}%
\allowdisplaybreaks
\usepackage{amsthm}%
\usepackage{mathrsfs}%
\usepackage[title]{appendix}%
\usepackage{xcolor}%
\usepackage{textcomp}%
\usepackage{manyfoot}%
\usepackage{booktabs}%
\usepackage{algorithm}%
\usepackage{algorithmicx}%
\usepackage{algpseudocode}%
\usepackage{listings}%

\usepackage{placeins}

\usepackage{natbib}
\usepackage{hyperref}
\hypersetup{
    colorlinks=true,
    linkcolor=blue,
    citecolor=red,
    urlcolor=blue
}

\usepackage{latexsym}
\usepackage{bm}
\usepackage{caption}
\usepackage{subcaption}

 \def\1{\raisebox{2pt}{\rm{$\chi$}}}

\numberwithin{equation}{section}

\newcommand{\R}{{\mathbb R}}

\newcommand{\D}{{\mathcal D}}
\newcommand{\E}{{\mathbb E\,}}

 \newcommand{\eps}{{\varepsilon}}
 
\newcommand{\abs}[1]{\left|#1\right|}
\newcommand{\norm}[1]{\left|\left|#1\right|\right|}

\def\vint_#1{\mathchoice%
          {\mathop{\kern 0.2em\vrule width 0.6em height 0.69678ex depth -0.58065ex
                  \kern -0.8em \intop}\nolimits_{\kern -0.4em#1}}%
          {\mathop{\kern 0.1em\vrule width 0.5em height 0.69678ex depth -0.60387ex
                  \kern -0.6em \intop}\nolimits_{#1}}%
          {\mathop{\kern 0.1em\vrule width 0.5em height 0.69678ex
              depth -0.60387ex
                  \kern -0.6em \intop}\nolimits_{#1}}%
          {\mathop{\kern 0.1em\vrule width 0.5em height 0.69678ex depth -0.60387ex
                  \kern -0.6em \intop}\nolimits_{#1}}}
\def\vintslides_#1{\mathchoice%
          {\mathop{\kern 0.1em\vrule width 0.5em height 0.697ex depth -0.581ex
                  \kern -0.6em \intop}\nolimits_{\kern -0.4em#1}}%
          {\mathop{\kern 0.1em\vrule width 0.3em height 0.697ex depth -0.604ex
                  \kern -0.4em \intop}\nolimits_{#1}}%
          {\mathop{\kern 0.1em\vrule width 0.3em height 0.697ex depth -0.604ex
                  \kern -0.4em \intop}\nolimits_{#1}}%
          {\mathop{\kern 0.1em\vrule width 0.3em height 0.697ex depth -0.604ex
                  \kern -0.4em \intop}\nolimits_{#1}}}

\newcommand{\aveint}[2]{\mathchoice%
          {\mathop{\kern 0.2em\vrule width 0.6em height 0.69678ex depth -0.58065ex
                  \kern -0.8em \intop}\nolimits_{\kern -0.45em#1}^{#2}}%
          {\mathop{\kern 0.1em\vrule width 0.5em height 0.69678ex depth -0.60387ex
                  \kern -0.6em \intop}\nolimits_{#1}^{#2}}%
          {\mathop{\kern 0.1em\vrule width 0.5em height 0.69678ex depth -0.60387ex
                  \kern -0.6em \intop}\nolimits_{#1}^{#2}}%
          {\mathop{\kern 0.1em\vrule width 0.5em height 0.69678ex depth -0.60387ex
                  \kern -0.6em \intop}\nolimits_{#1}^{#2}}}
\newcommand{\ud}{\, d}
\newcommand{\half}{{\frac{1}{2}}}

\newcommand{\tr}{\operatorname{tr}}

\theoremstyle{definition}%
\newtheorem{lemma}{Lemma}

\theoremstyle{remark}%
\newtheorem{remark}{Remark}%

\begin{document}

\title{Dynamic programming principle in cost-efficient sequential design:    optimal update scheduling under cost constraints}



\author{
Jeongmin Han\thanks{Email: jeongmin.han@ssu.ac.kr}\\
\affil{Soongsil University}
}

\author{
\name{Jeongmin Han\textsuperscript{a}\thanks{CONTACT Jeongmin Han. Email: jeongmin.han@ssu.ac.kr}, Juha Karvanen\textsuperscript{b} and Mikko Parviainen\textsuperscript{b}}
\affil{\textsuperscript{a}Department of Mathematics, Soongsil University, Seoul, Republic of Korea; \textsuperscript{b}Department of Mathematics and Statistics, University of Jyv\"askyl\"a, Jyv\"askyl\"a, Finland}
}

\maketitle



\abstract{We study sequential cost-efficient design in a situation where each update of covariates involves a fixed time cost typically considerable compared to a single measurement time.   
The problem arises from parameter estimation in switching measurements on superconducting Josephson junctions which are components needed in quantum computers and other superconducting electronics.
In switching measurements, a sequence of current pulses is applied to the junction and a binary voltage response is observed. The measurement requires a very low temperature that can be kept stable only for a relatively short time, and therefore it is essential to use an efficient design. 
We use the dynamic programming principle from the mathematical theory of optimal control to solve the optimal update times.   Specifically, we give a formulation for $D$-optimal experimental design based on the dynamic programming principle with a binary response model.
 
Our simulations demonstrate the cost-efficiency compared to the previously used methods.}

\keywords{optimal design, $D$-optimality, complementary log-log link, Josephson junction.}



\section{Introduction}
\label{sec:intro}

\noindent
\leftskip=0pt
\rightskip=0pt

We study an optimal design problem motivated by switching measurements in quantum physics.
  Before using a Josephson junction (JJ) \citep{josephson62}, an important nonlinear component in superconducting electronics utilized for example in quantum computers \citep{makhlinss01}, the parameters of the particular component need to be determined by measurements.
In these measurements, a component is placed at a temperature close to absolute zero to obtain superconductivity. When a current pulse is applied to the junction, a voltage pulse is observed with a probability that depends on the height of the current pulse. The relation between the height of the current pulse and the probability of the voltage pulse can be accurately approximated by a binary response model with a complementary log-log link \citep{karvanenvtp07}. The unknown model parameters $a$ and $b$ determine the slope and the location of the response curve and depend on the physical dimensions of the junction. 
The problem is to choose the covariate values, i.e., the height of the applied current pulse in our case, in order to estimate the parameters of interest in the most efficient manner in the sense of $D$-optimality (see Section~\ref{sec:D-optimal}). Efficiency is essential because in switching measurements, the required temperature can be kept stable only for a relatively short time.

 Sequential experimentation provides benefits over static designs. In nonlinear models, the optimal design depends on the parameters. As the true values of the parameters are unknown, the optimality properties of a design depend on the accuracy of initial estimates of the parameters. A sequential design allows us to update the estimates after each new observation and thus improve the design. 
In practice, sequential experimentation has its price. In switching measurements, adjusting the pulse generator to change the covariate values takes a time that is considerably larger than the time needed for a single measurement. 
This leads to cost-efficiency considerations where one must balance the cost of an update and the information obtained from it. The sequential design is then replaced by a batch-sequential (group-sequential) design, and the key problem is determining the optimal update times. 

The batch-sequential design problem in the context of switching measurements was formulated by \citet{karvanenvtp07}. They implemented an ad hoc method that increased the number of measurements by 10 \% at each stage. Later by \citet{karvanen09}, an approximate asymptotic model for the accumulation of a quantity approximating the $D$-criterion was presented. Moreover, an ad hoc method for solving the optimal update times was implemented. The purpose of the current paper is in the context of the approximate asymptotic model to solve the optimal update times in a precise manner using the mathematical theory of optimal control, and the dynamic programming principle (DPP) in particular.

\citet{huanm} proposed a DPP approach for optimal sequential
experimental design accounting for future information obtained from subsequent
experiments. They demonstrated the method through Bayesian simulations of optimal
design problems with continuous responses under linear or nonlinear regression
models using appropriately specified reward functions. In contrast, our study is the first to
formulate a DPP based experimental design framework for a binary response model.
Moreover, while \citet{huanm} focused on dynamic programming
problems with a fixed number of stages, our work also accommodates situations in which the
number of stages is not predetermined.  

 The DPP roughly speaking means that instead of considering a process until the end we look at the problem over one step or stage. A simple example is given by a random walk where we jump left or right each with probability $1/2$ on the integers $\{1,2,\ldots,9\}$ and stop if we reach $0$ or $10$ where we get some given payoff. When looking at the expectation in this process when starting at $x_0$ (denoted by $u(x_0)$), the  DPP  for the expectation reads as
\begin{align*}
    u(x_0)=\half\big(u(x_0-1)+u(x_0+1)\big).
\end{align*}
In other words, this says that the expectation at a point can be considered by looking at one step and summing up the outcomes with the corresponding probabilities: we either go left with probability $1/2$ or right with probability $1/2$ and take the expectations for the function value.
More examples can be found for example in \citet[Section 10.3]{evans10}, \citet{bardic97} or \citet[Chapter 1]{parviainenb}. 
In the design of experiments, dynamic programming has been applied to sequential experimentation with linear models in \citet{ben-galc02,huanm}. 
We also refer to \citet{murphy2003,gittinsjones,villar18}, which are concerned with applications of dynamic programming problems to designs of clinical trials.

The key dynamic programming equations in our case are formulated in Sections \ref{sec:maximizeD} and \ref{sec:minimizetime}. From these, we then solve the optimal update times using R \citep{R23}. In particular, we take into account which target (time or optimal $D$) we are optimizing, and present the method to solve both of these problems. Moreover, the dynamic programming principles automatically take into account the amount of information we have obtained during the initialization and the update cost when determining the optimal update times.

In Section \ref{sec:benchmark}, we present the results obtained through a simulation testing the DPP method and previous methods in the literature. The results show that the suggested method is more cost efficient. Finally, we remark that the use of optimal control should find a wide variety of applications in cost-efficient sequential designs with different statistical models. It is by no means limited to the presented examples. 
\section{Preliminaries: $D$-optimal designs for complementary log-log regression}

\label{sec:D-optimal}

In this paper, we deal with the DPP in the context of $D$-optimal designs, where $D$ is the determinant, or equivalently, the square root of the determinant of the information matrix. $D$-optimality is a popular criterion that maximizes the determinant of the information matrix. It does not rely on any prior knowledge about the parameters being estimated. This makes it particularly useful in situations where little or no information is available about the parameters like in the Josephson junction example.

The preliminaries of the DPP are given in the appendix and the next section: before we can write down the DPP, we need to recall the basic notation for the $D$-optimal design.  The process of deriving the (locally) $D$-optimal design involves maximizing the criterion $D$ under certain assumptions. A detailed exposition of this derivation in a generalized form is presented in \citet{ford92}. Moreover, the extensively employed logit model has been explored in various works, including studies by \citet{abdelbasitp83}, \citet{minkins87}, \citet{sitterw93}, as well as \citet{mathews01}. These references collectively demonstrate that the $D$-optimal design represents a two-point design applicable to logit, probit, and cloglog models, enabling the determination of optimal covariate values. Additionally, extensions addressing models involving multiple covariates have been investigated by \citet{woodsler06}, \citet{drors08}, and \citet{dortaguerragg08}.

Here we consider an experiment, where our task is to estimate the unknown parameters $\boldsymbol \theta=(a,b)$. We assume that the binary random variable is distributed according to  
\begin{align}
\label{eq:probs}
&P(Y=1|\boldsymbol \theta)=1-e^{-\exp(ax+b)},\nonumber \\
&P(Y=0|\boldsymbol \theta)=e^{-\exp(ax+b)},
\end{align}
where $x$ is the covariate. 

The task is to select the covariate values $x_1,\ldots,x_n$ in such a way that the parameters can be estimated from the observed data $y_1, \ldots,y_n$ as efficiently as possible.  
 From (\ref{eq:probs}), the likelihood of $n$ iid observations is represented by 
\begin{align*}  
L( \boldsymbol \theta | \bm y)   &:=
\prod_{i=1}^n P(Y=y_i| \boldsymbol \theta)\\
&:=\prod_{i=1}^n\Big\{y_i(1-e^{-\exp(ax_i+b)})+(1-y_i)e^{-\exp(ax_i+b)}\Big\}.
\end{align*}
Thus the log-likelihood is
\begin{align}
\label{eq:log-likelihood}
l(\boldsymbol \theta):
&= \sum_{i=1}^n \log P(Y=y_i|\boldsymbol \theta)\nonumber \\
&=\sum_{i=1}^n\Big\{y_i\log((1-e^{-\exp(ax_i+b)}))-(1-y_i)\exp(ax_i+b)\Big\}.
\end{align}

The Fisher information matrix is given by 
\begin{align*}
J=-\E\begin{pmatrix}
\frac{\partial^2l}{\partial a^2}&\frac{\partial^2l}{\partial a\partial b}\\
\frac{\partial^2l}{\partial b\partial a}& \frac{\partial^2l}{\partial b^2}
\end{pmatrix}=\begin{pmatrix}
\sum_{i=1}^ng(z_i)x_i^2&\sum_{i=1}^ng(z_i)x_i\\
\sum_{i=1}^ng(z_i)x_i & \sum_{i=1}^ng(z_i)
\end{pmatrix},
\end{align*}
where $z_i=ax_i+b$ and
\begin{align*}
g(z)=\frac{e^{2z}}{e^{\exp(z)}-1}.
\end{align*}
Observe that the Fisher information depends on the sample size but not on the measurements $y_i$. Heuristically speaking, the Fisher information measures the curvature of the graph of log-likelihood and thus tells how stable the maximum likelihood estimate is. 

We formulate the $D$-criterion in terms of
\begin{align*}
    D:=\sqrt{\det(J)}.
\end{align*}
In $D$-optimal design, our task is to find $x_1,\ldots,x_n$
so that $D$ (or equivalently $D^2$) is maximized. 
Here, we selected the square root instead of the determinant in order to have correct scaling properties later on. By reorganizing the terms, we obtain
\begin{align*}
D^2:=\det(J)&=\sum_{i=1}^ng(z_i)x_i^2\sum_{i=1}^ng(z_i)-\Big(  \sum_{i=1}^ng(z_i)x_i\Big)^2\\
&=\sum_{i=1}^{n-1} \sum_{j=i+1}^{n} g(z_i)g(z_j)(x_i- x_j)^2.
\end{align*}
The $D$-optimal design is a two-point design and we can maximize 
\begin{align} \label{twoptdes}
    D^2=\frac{1}{a^2}g(z_1)g(z_2)(z_1-z_2)^2
\end{align}
to obtain the values
\begin{align*}
(z_{1}^{*},z_{2}^{*})=(0.97963269129,-1.337736677)
\end{align*} 
as well as $(z_{1}^{*},z_{2}^{*})=(-1.337736677,0.97963269129)$ (see \citet{ford92}). The Fisher information matrix corresponding to these values is denoted by $J^*$. In other words, $J^*$ is the information matrix for $D$-optimal covariates when $n=2$.

In order to solve the optimal update times later, we need to model how much information we will have after an additional measurement given the so far acquired information. We use the approximation from \citet{karvanen09} for the expected change in $D$ when two additional measurements are made.  Our intention here is not to study the model there, but rather to show, given such a model, how to solve the optimal update times using the dynamic programming principle. In any case, we briefly recall the setting in  \citet{karvanen09}. 
Similarly, for now, we assume that $(a,b)=(1,0)$ and only comment on the other cases later. We set
\begin{align*}
J^{*}&=\sum_{i=1}^{2} g(z^{*}_{i})
\begin{pmatrix}
(x^{*}_i)^2&x^{*}_i\\
x^{*}_i&1 
\end{pmatrix},
\end{align*}
where $z^{*}_i=ax^{*}_i+b=x^{*}_i$, and observe that $\sqrt{\det(J^{*})}=0.80940268$.  
Motivated by the asymptotic normality of maximum likelihood estimates it was assumed that
$\hat a$ and $\hat b$ are normally distributed with the mean $(a,b)$ and with the covariance matrix $J^{-1}(D)$, where $J(D)$ represents the accumulated information. 
Then in the simulation pairs  $(\hat a, \hat b)$ were generated from the distribution $N((1,0),J^{-1}(D))$, and for each simulated $(\hat a,\hat b)$, the following quantity was computed:
\begin{align*}
\det\Bigg(\sum_{i=1}^{2} g(z_i)
\begin{pmatrix}
x_i^2&x_i\\
x_i&1 
\end{pmatrix}\Bigg)^{1/2}
&=\sqrt{g(z_1)g(z_2)}(z_1-z_2),
\end{align*}
where
\begin{align*}
z_1 &=\frac{z_{1}^{*}-\hat b}{\hat a},\quad \quad z_2 =\frac{z_{2}^{*}-\hat b}{\hat a}.
\end{align*} 
The expected change in $D$ was taken to be the average of these simulated quantities,  
which is denoted by the quantity $h(D)$.  
More details are given in Appendix~\ref{sec:acc-D}. 
\begin{figure}
  \centering
  \includegraphics[width=0.8\linewidth]{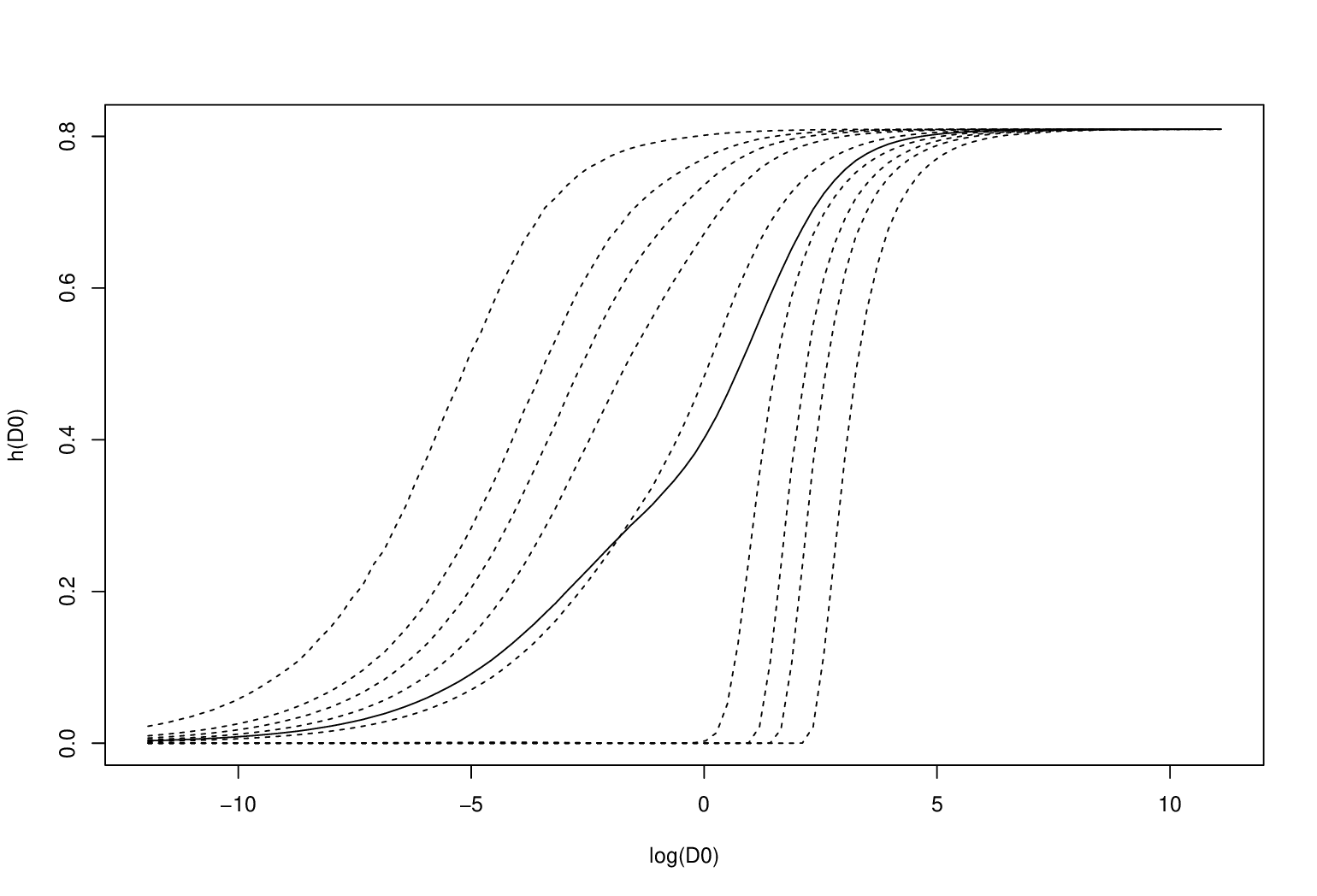}
\caption{The above figure represents the simulation result of the expected information increase $h$ in \citet{karvanen09}.
The solid line indicates the average of $h(D)$;
dotted lines indicate 1st, 5th, 10th, 20th, 50th, 80th, 90th, 95th and 99th percentiles of $h(D)$ in order from left.}
\label{fig:1}
\end{figure}
In \citet{karvanen09}, see Figure~\ref{fig:1}, the following logistic model
 \begin{align}
 \label{eq:h}
h(D)=\frac{0.80940268}{1+e^{\eta+\theta\log(D)}}
\end{align}
 with
  $\eta=-0.66081,\ \theta= -1.24852$,
was suggested as an approximation of the expected change in $D$ when another measurement is being made.
We also remark that $\lim_{D\to0+} h(D)=0$. 
Therefore, we also need an initialization procedure that gives us the initial $D_0>0$ and the first MLE to get started.
 Meanwhile, for the function $h$, we may also consider alternative approximations; in such cases, the method we present in the next section still works well (see the defintion of $D_{\text{next}}$ in  \eqref{eq:dpp-discr}).

Now we consider the situation where covariate updates are performed at times $t_0, \dots, t_n$.
By definition, the function $h$ represents the expected change in $D$ under the two-point design. 
If measurements are conducted without updating the covariates, the same $x_i^{\ast}$ values are repeatedly used. 
Based on the approximation in \eqref{approx}, the expected change in $D$ would also remain constant until a new update occurs. Hence, the slope of $D$ in the model, $h(D(t_i))$, remains constant over the interval $[t_i,t_{i+1}]$ (see \eqref{eq:dpp-discr}).

\FloatBarrier

\section{Optimal update times}

\label{sec:dpp-opt-times}
Given the model described in the previous section, we now show how to solve optimal update times using dynamic programming. We consider two problem settings: given the available time, we want to maximize $D$, or given the desired $D$, we want to minimize the time used to reach this target. The dynamic programming principles below can easily be solved numerically by iterating backwards from the final condition. 

\subsection{Given final time, maximize $D$} 
\label{sec:maximizeD}

We consider a situation where the termination time $T$ is given, and we want to maximize $D$. For example, it is known that the Josephson junction remains at a required temperature and thus is stable for a certain time $T$. The time cost $C_s$ for updating the covariate is assumed to be a nonnegative integer and is here usually large compared to the cost of a single measurement (the cost of the single measurement is scaled to be 1). For example, $C_s$ could be the time cost for adjusting the current in the pulse generator. The task is to find the optimal update times so as to accumulate as large $D$ as possible by the termination time $T$. 
The problem is to decide the next update time 
at the current update time $t$, assuming that $D\ge D_0>0$ is given. Let us denote by $u(D,t)$ the amount of $D$ we can optimally accumulate during the rest of the experiment if starting at $(D,t)$ and updating immediately.  
Naturally, we do not at this point know how exactly the updates and measurements should be made later on in order to optimally accumulate $D$ in the measurements, but this is exactly what we will solve from the DPP. This description is enough for now, but the precise definition is given in Appendix \ref{sec:math}. Using the terminology from the control theory, the value function $u$ satisfies the DPP with $C_s>0$ and $t<T-C_s$  
\begin{align}
\label{eq:dpp-discr}
 u(D,t)=\sup_{t_{\text{next}}\in (t+C_s,T]}
\Big\{ u(D_{\text{next}},t_{\text{next}})+(D_{\text{next}}-D)\Big\},
\end{align}
where
\begin{align*}
D_{\text{next}}:&=D_{\text{next}}(t_{\text{next}},D):=D+h(D)(t_{\text{next}}-t-C_s),
\end{align*} and
\begin{align*}
u(D,t)&=0,\quad t\in [T-C_s,T]. 
\end{align*}  
The DPP \eqref{eq:dpp-discr} can be regarded as being derived from \eqref{approx} in Appendix A. The estimate \eqref{approx} in the Appendix \ref{sec:acc-D} shows that the information $D$ increases approximately linearly as additional measurements are collected when the sample size is sufficiently large. Motivated by this approximation, we constructed \eqref{eq:dpp-discr} so that $D$ increases linearly over time.  
We also remark that the solution to this DPP is unique and stable (see Appendix \ref{sec:math}).  
We also note here that what the DPP says is that we can compute the optimal value at a point by taking one optimal stage of measurements and the optimal value from where we ended up during this one stage. 
We note that in our simulation, the variables $D$ and $t$ are both discretized in integer units. Here, $t_{\text{next}}$ is selected from among the integer values of $t$ that satisfy the given condition, and $D_{\text{next}}$ is computed for each integer value of $D$.

From \eqref{twoptdes}, we recall that
$$D^2 =\frac{1}{a^2}g(z_1)g(z_2)(z_1-z_2)^2$$
for a two-point design, which is the quantity to be maximized.
 
Then we observe that it does not matter that it was assumed that $a=1$ in $h$ since everything would be just scaled by the true value of $a$, and this does not affect the update times. However, inside the DPP, the quantity $D$ corresponds to $a=1$ (we denote it by $D_{\text{model}}$) and should not be taken as an approximation of the 'true $D$', which will be denoted by $D_{\text{accum}}$. Nonetheless, $D_{\text{accum}}$ can easily be computed from the sample.  

\begin{figure}[!htbp] 
\centering
\begin{subfigure}{.4\textwidth}
  \centering
  \includegraphics[width=1\linewidth]{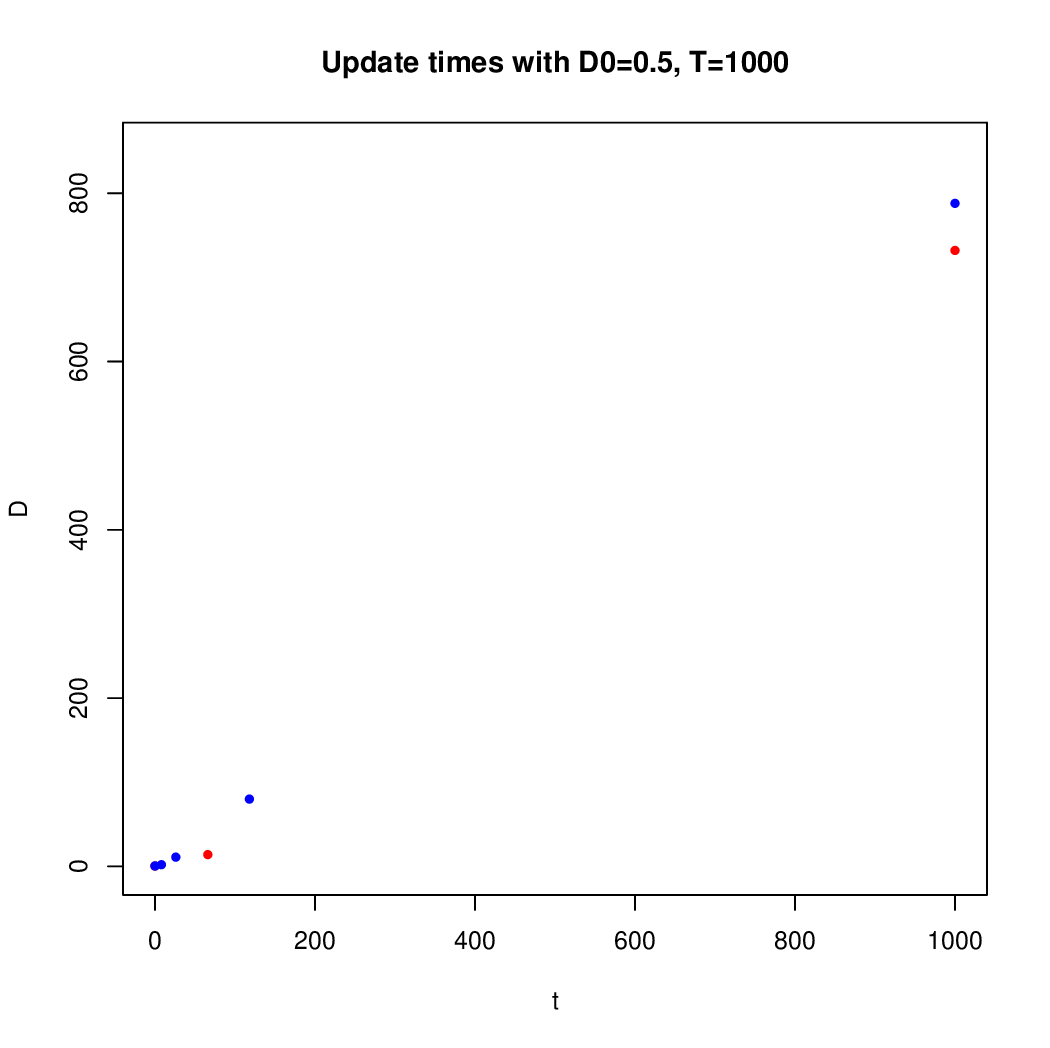}
  \caption{Changing $C_s$ ($D_0=0.5,\ T=1000$)}
  \label{fig:sub1}
\end{subfigure}%
\begin{subfigure}{.4\textwidth}
  \centering
  \includegraphics[width=1\linewidth]{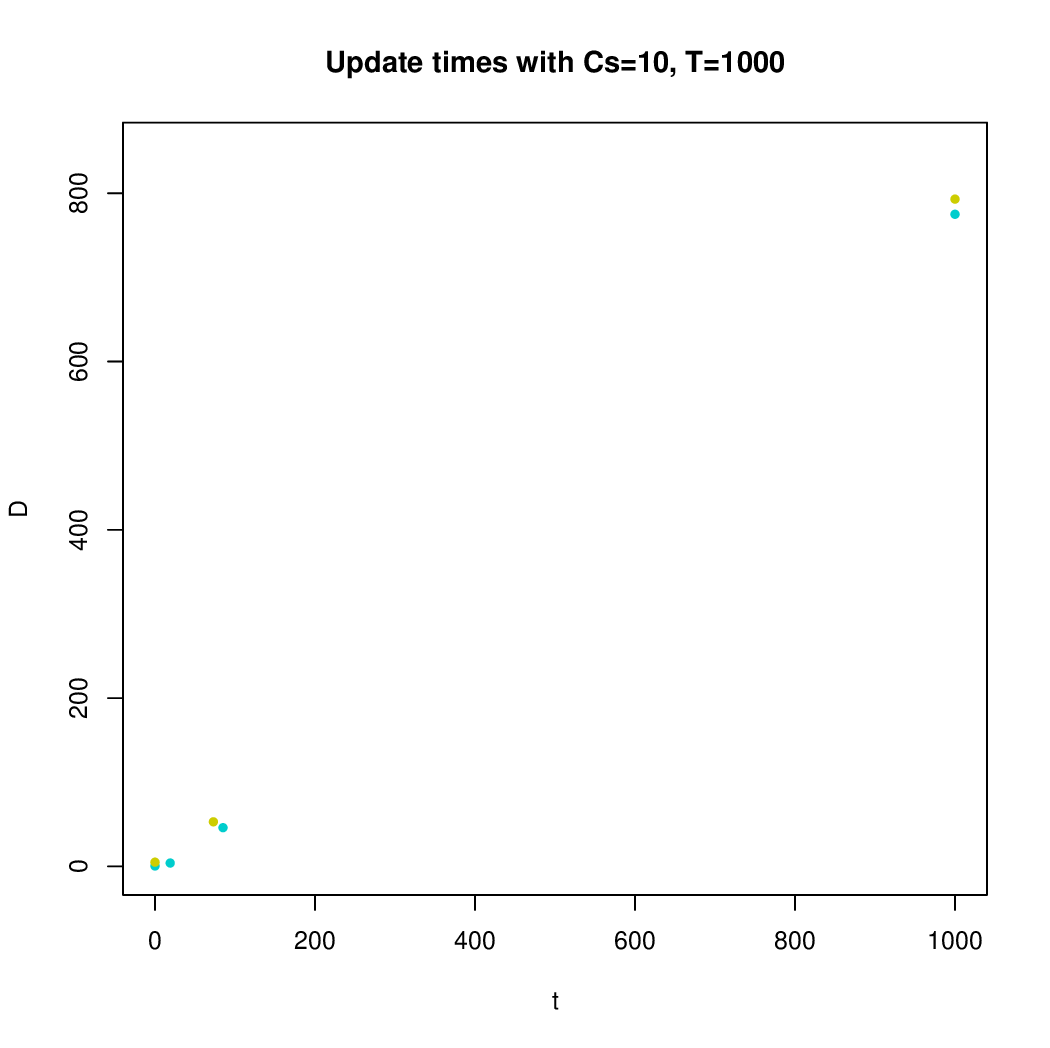}
  \caption{Changing $D_0$  ($C_s=10, \ T=1000$)}
  \label{fig:sub2}
\end{subfigure}
\begin{subfigure}{.4\textwidth}
  \centering
  \includegraphics[width=1\linewidth]{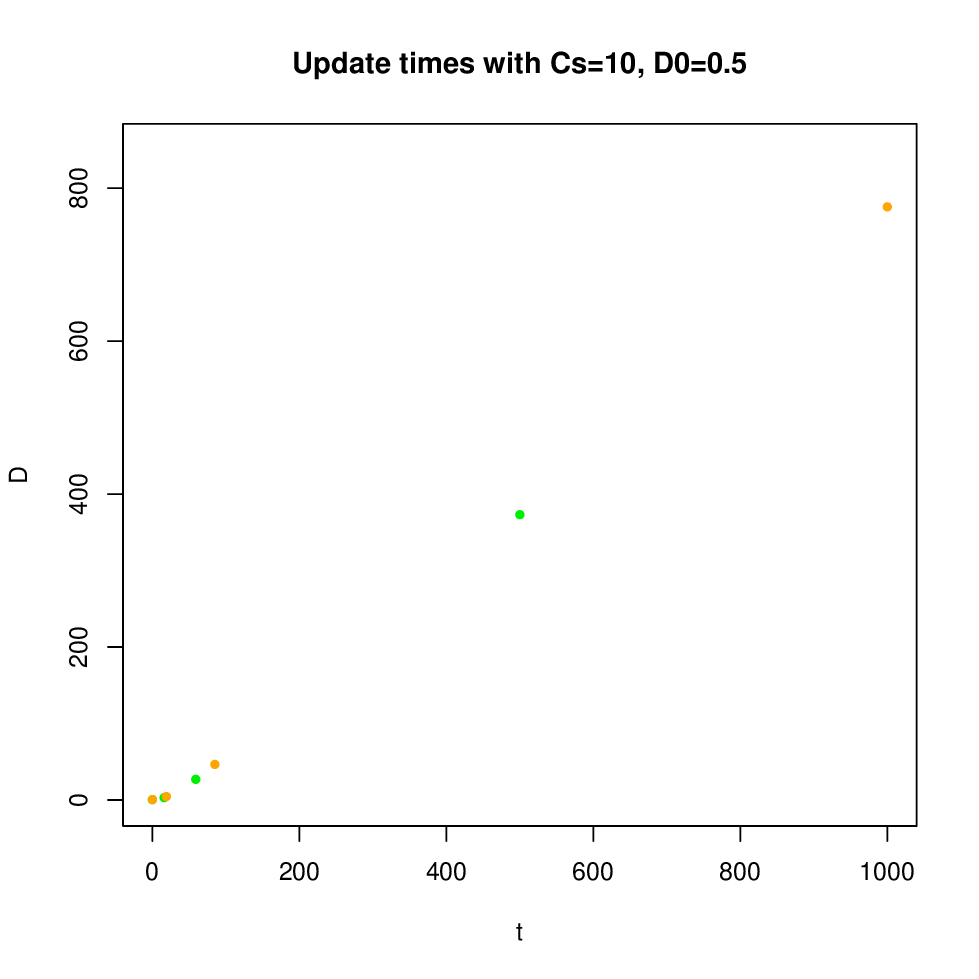}
  \caption{Changing $T$  ($C_s=10, \ D_0=0.5$)}
  \label{fig:sub3}
\end{subfigure}
\caption{There are three examples of update times (the termination time $T$ is not an update time but shows the final $D$). The first subfigure indicates update times for fixed $D_0=0.5$ and $T=1000$ ($C_s=5$ in blue, $C_s=30$ in red). The second one represents update times for fixed $C_s=10$ and $T=1000$ ($D_0=0.5$ in cyan, $D_0=5$ in yellow). The last subfigure shows the simulation for fixed $C_s=10$ and $D_0=0.5$ ($T=500$ in green, $T=1000$ in orange). We can observe that larger values of $C_s$, $D_0$ or $T$
delay the occurrence of the first update, thereby subsequently affecting the points and the frequency of updates.
When the initial parameter estimates are obtained differently, the corresponding $D_0$ value changes, and accordingly, the optimal update time is also computed differently.}
\label{fig:times}
\end{figure}

In practice, solving this DPP requires a backward iteration based on the assumption that $u(D,t) = 0$ for $t \in [T-C_s, T]$.
Specifically, since $t$ is discretized in integer units, $u(D,t)$ for each integer $D$ can be computed sequentially from the largest integer $t$ smaller than $T-C_s$ down to $0$, by comparing $ u(D_{\text{next}},t_{\text{next}})+(D_{\text{next}}-D)$.
We will solve for $u$ starting from $t=T-C_s-1$. When at $(D,t)$, supposing that we have already solved values $u$ for all $s>t$, we may solve for $u(D,t)$, using (\ref{eq:dpp-discr}).
In the course of using (\ref{eq:dpp-discr}), we also solve and save the stage update times for each $D$. Then we pick the stage update time $t_{next}$ for $(D,t)=(D_0,0)$, and by continuing, for the rest of the stages. 
Figure \ref{fig:times} gives three examples.

The optimization is implemented in the R codes available at optimaltimesCsA.r, optimaltimesCsB.r and optimaltimesCsC.r,
which are publicly available at \href{https://github.com/JeongminHan-cloud/OptimalSequentialDesign}{https://github.com/JeongminHan-cloud/OptimalSequentialDesign}.  
In these R-codes, for example, optimaltimesCs.r, we have discrete grids for $D_{i},\ i=1,\ldots,N_D$ and $t_{j},\ j=1,\ldots, N_T$. When at a point $(D_{i_0},t_{j_0})$, we consider the time points $\tilde t_j\in\{\tilde t_{1},\ldots, \tilde t_{N_{t_{j_0}}}\}:= \{t_{j_0}+C_s,\ldots, t_{N_T}=T\}$ and compute the values 
$$
\tilde D_{\tilde t_j}:=D_k+h(D_k)(\tilde t_{j}-t_{j_0}-C_s).
$$
 On the other hand, we have already computed all the future values for $u$ at the discretized values of $D_{i},\ i=1,\ldots,N_D$. Thus rounding  $\tilde D_{t_j}$ to the nearest grid value $D_{k}$, we know $u(D_{k},\tilde t_{j})$ and we may form 
\begin{align*}
DPP_{i_0,j}:=u(D_{k},\tilde t_{j})+(\tilde D_{\tilde t_j}-D_{i_0}),\quad \text{for all} \quad j=1,\ldots, N_{t_{j_0}}.
\end{align*}
Now, according to (\ref{eq:dpp-discr}), we have solved the values of $u$ at time $t_{j_0}$. Then we continue to time $t_{j_0}-1$.

\FloatBarrier

\subsection{Given desired $D_{\text{final}}$, minimize the time cost}
\label{sec:minimizetime}
In this section, we consider a different problem where the goal is to reach a given final value of accuracy $D_{\text{final}}$ in the shortest possible time.  
Let us denote by $v(D)$ the minimal time to reach $D_{\text{final}}$ if we update immediately.
The experimenter can choose $D_{\text{final}}$, for instance, by setting a target for standard errors of the model parameters and calculating the corresponding $D$-criterion. 
Again, for a more detailed discussion, see Appendix \ref{sec:math}. First, as indicated above, we observe that 
$v$ does not depend on the time spent so far since we only look at the future accumulation of $D$.
We still denote the change of $D$ in one double measurement by $h(D)$ given in (\ref{eq:h}), and by $C_s$ the update cost of the covariates.
Now $v$ satisfies the following DPP
\begin{align}
\label{eq:dppt-discr}
v(D)&=\inf_{\Delta t\in (C_s,\infty)}
\Big\{ v\big(D+(\Delta t-C_s)h(D)\big)+\Delta t\Big\},\\
v(D)&=0\quad \text{ for }\quad D\ge D_{\text{final}}.\nonumber
\end{align}
Recall that we only applied $(a,b)=(1,0)$ in the previous subsection,
since the accumulated information in the general case can be obtained by scaling everything by the factor $a^{-2}$ with the true value of $a$. One should observe that also in the above DPP $D$ and $D_{\text{final}}$ denote the rescaled versions with $(a,b)=(1,0)$ and should not be taken as an approximation of $D_{\text{accum}}$. Also, the updates should not be launched by the observed $D_{\text{accum}}$ computed from the sample but by the time steps $\Delta t$ that will be stored while numerically solving for $v$.

To finish the section, observe that the optimal update times depend (as they should) on whether we are minimizing the time cost or maximizing $D$, what are the final values of $T$ or $D_{\text{final}}$, the update time cost $C_s$, and $D_0$ after the initialization. Thus it is not practical to give update times in a table, but they can be solved for different values by using the provided codes.

\section{Benchmark}
\label{sec:benchmark}

 In this section, we present the simulation results for the cost-efficient sequential design for switching measurements in superconducting Josephson junctions. As explained in Section \ref{sec:dpp-opt-times}, the code optimaltimesCs.r calculates the optimal update times for the covariates using the DPP method in the case of a given final time $T$. The aim is to estimate the parameters in the model as precisely as possible in this given time frame. Typically, it is known that the Josephson junction only remains in the superconducting temperature for the time $T$.  In this particular example, the covariate is the current pulse height but the methodology works in great generality. The point here is that adjusting the pulse generator for optimal pulse height (in the sense of $D$-optimality in this particular example) for re-estimated parameters is slow compared to a single measurement, and thus one can predict that the optimal solution is at least not to update covariates after every measurement. Then in simulation.r  (or simulation\texttt{\_}largeD.r), we test the method by simulating the measurements, keeping track of the accumulated data. The codes together with more precise descriptions in readme.txt are also provided.

 We give a brief overview of the simulation conducted in the code simulation.r here.  
In preceding studies \citep{karvanenvtp07,karvanen09}, several types of batch-sequential $D$-optimal design methods were presented.
They used the measurement data for a JJ circuit given in 
\href{https://github.com/JuhaKarvanen/switching_measurements_design}
{https://github.com/JuhaKarvanen/switching\_measurements\_design}.  
In addition to the existing methods, we simulate the experiment with our suggested DPP-based method. 
The method based on a regression model used in \citet{karvanen09} provides approximately cost-efficient update timings in many cases. 
However, when the initial $D$-value $D_0$ is too small, it may cause excessive updates in the early stages of the experiment, leading to increased costs.
A small initial value of $D_0$ indicates that the initial parameter estimates are highly inaccurate. Since such inaccurate initial estimates are quite common in many experimental settings, it is meaningful to consider alternative methodologies that offer improved cost efficiency under these circumstances.

\begin{figure}[!htbp]
  \centering
  \includegraphics[width=0.65\linewidth]{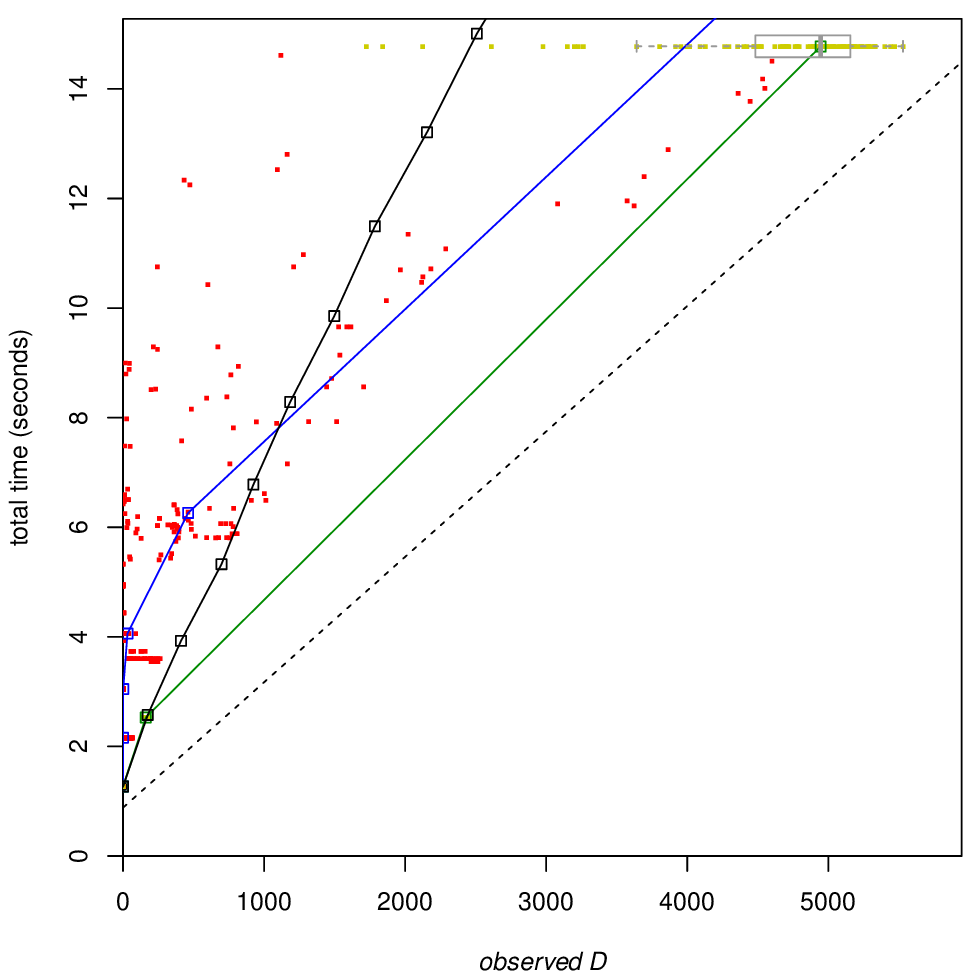}
  \label{fig:sub3}
\caption{Simulations for three methods with $D_0=0.1408$ and $(a,b)=(0.24,-61)$.  The black line marks the simulated results obtained in the same way as the ad hoc method used in the original experiment in \citet{karvanenvtp07}. The blue line and green line represent the time costs obtained by the approximate cost-efficient method in \citet{karvanen09} and the DPP method of this paper, respectively. The blue and cyan squares represent the path, and the corresponding experiment design is calculated as the median of $D_{\text{accum}}$ of 100 simulated paths.
The dots show the stages of the 100 simulated paths (red: design in \citet{karvanen09}, yellow: DPP design).
The dashed line indicates the theoretical limit of the time cost if they are immediately measured at the optimal points.
In the DPP design, we set $T=3500$ and the total time cost is 14.77313 sec in this case.
Our objective is to achieve the same value of $D$ with minimal consumed time (cost); thus, the proximity of a graph to the dotted line reflects higher efficiency of the experimental design. Among the three, the DPP-based design demonstrates the greatest efficiency; compared with \citet{karvanen09}, the time cost was reduced by 13.58\%.}
\label{fig:times2}
\end{figure}

We use less intense initialization with bad initial guesses of $a$ and $b$ to obtain a lower value for $D_0$ and it can be reproduced through the code initialdatagen.r for the case when the stage size is 100 in the first round under the assumption that the true values of the parameters are  $(a,b)=(0.24,-61)$. This is in order to illustrate the difference of the methods more clearly (the lower $D_0$, the more important the updates as well as cost-efficient designs are).  
In the code simulation.r, we first solved the DPP problem \eqref{eq:dpp-discr} (given final time, optimize $D_{\text{model}}$ with $T=3500$ and $C_S=0.88167/0.00386\approx228.4$) and then obtained the optimal update times under our setting.
After that, we ran 100 simulations based on those update times.
We also conducted the same number of simulations for the model suggested in \citet{karvanen09}  with the target $D$-value set to 100000 for comparison.  
The results are illustrated in Figure \ref{fig:times2}.  As a technical remark, $D$ in the figure is the actual observed the accumulated $D$ ($D_{\text{accum}}$) computed from the sample, not $D_{\text{model}}$ that appears for example in the DPPs and is obtained using $h(D)$ and rescaling $(a,b)=(1,0)$. What we observe is that the DPP method is more cost-efficient than the model in \citep{karvanen09}.

\begin{table}[!htbp]  
\centering
\begin{tabular}{|c|c|c|c|}
\hline
\textbf{Stage} & \textbf{n} & $D_{\text{accum}}$ & std($D_{\text{accum}}$)   \\
\hline
1  & 100   & 0.14    &  -         \\
2  & 98   & 161.54    &   9.47           \\
3  & 2944   & 4945.63  & 762.78  \\
\hline
\end{tabular}
\caption{Accumulated $D$ ($D_{\text{accum}}$) with standard deviations across stages}
\label{tab:parameter_estimates}
\end{table}

\FloatBarrier

\section{Discussion}

We considered a cost-efficient batch-sequential design in a situation where the update cost is high compared to the cost of a single measurement. The work was motivated by switching measurements of Josephson junctions. 
 The preceding studies \citep{karvanenvtp07,karvanen09} have explored how to update parameter estimates in experiments observing this physical phenomenon in terms of cost-efficiency.  
We made several improvements to the earlier literature: We formulated and solved a precise dynamic programming problem for update times if the change in $D$ is given by the function $h(D)$ obtained from the literature. In particular, the dynamic programming approach automatically takes into account which target (target time or target $D$) we are optimizing, and the effect of $D_0$, $C_S$ on update times. 
The simulations illustrate the efficiency of the proposed approach. Compared to earlier approximations \citep{karvanenvtp07,karvanen09}, a clear gain in cost-efficiency was seen.

However, there are still several approximations stemming from the earlier literature. The particular assumed form of the Fisher information in the simulation producing $h$, the normality assumption of MLE, and the assumed accumulation property of $D$, are such approximations and only hold asymptotically.  The reason for these approximations is of course a desire to formulate the problem in such a way that it can be simulated and solved by efficient tools and in advance. 

The proposed approach is directly applicable to switching measurements. In addition, a similar approach could be useful in other applications where the data are collected in batches and a cost is associated with each update. The functions $h(D)$ have also been estimated for the logit and probit link functions \citep{karvanen09}. One might also want to consider the DPP approach with other optimality criteria.

\begin{appendices}
\section{Accumulation of $D$ in the approximative model}
\label{sec:acc-D}

To obtain an approximative model for the accumulation of $D$, the expected change in $D$ in \citet{karvanen09} was taken to be the average of the simulated quantities $\sqrt{\det(J_i)}$,  where $J_i=J_i(\omega)$ denotes the information obtained in the round $i$ for each sample $\omega$ (we use the notation $J_i$ for the convenience).  Since the model is founded on such an accumulation property that holds asymptotically (similarly to the previous steps in the model), we give a brief justification for it here. 
To be more precise, we want to show in a suitable mode of stochastic convergence that 
\begin{align} \label{approx}
\sqrt{\det(J_1+\ldots+J_n)}- \sqrt{\det(J_1+\ldots+J_{n-1})}- \sqrt{\det(J_n)}\to0
\end{align}
almost surely as $n\to\infty$.  In other words, the difference between the first two terms reflects the actual change in $D$ if we obtain the information $J_n$, and the last term the quantity used in the simulation.  
To establish the above asymptotic behaviour, let us denote
$\Delta J_i:=(J_i-J^*)$ and compute
\begin{align*}
&\sqrt{\det(J_1+\ldots+J_n)}=\det\Big(nJ^*+\sum_{i=1}^n \Delta J_i\Big)^{1/2}\\
&=n\det\Big(J^*+\frac1n\sum_{i=1}^n\Delta J_i\Big)^{1/2}\\
&=(n-1)\det\Big(J^*+\frac1n\sum_{i=1}^n\Delta J_i\Big)^{1/2}+\det\Big(J^*+\frac1n\sum_{i=1}^n\Delta J_i\Big)^{1/2}.
\end{align*}
To get an approximation of 
$$ \sqrt{\det(J_1+\ldots+J_n)}- \sqrt{\det(J_1+\ldots+J_{n-1})},$$ in order to prove (\ref{approx}) we also observe that 
\begin{align*}
    & \sqrt{\det(J_1+\ldots+J_{n-1})}=(n-1)\det\Big(J^*+\frac1{n-1}\sum_{i=1}^{n-1}\Delta J_i\Big)^{1/2}
   \\
   &
=   (n-1) \det\Big(J^*+\frac1{n}\sum_{i=1}^{n}\Delta J_i\\
&\hspace{7 em}+\frac1{n-1}\Big(\sum_{i=1}^{n-1}\Delta J_i-\sum_{i=1}^{n}\Delta J_i\Big)+\Big(\frac1{n-1}-\frac1{n}\Big)\sum_{i=1}^{n}\Delta J_i\Big)^{1/2}
 \\
   &
=   (n-1) \det\Big(J^*+\frac1{n}\sum_{i=1}^{n}\Delta J_i-\frac1{n-1}\Delta J_n+\frac1{n(n-1)}\sum_{i=1}^{n}\Delta J_i\Big)^{1/2}\\
&=:(n-1)\det(A+B)^{1/2},
\end{align*}
where the shorthand notation $A$ contains the first two terms and $B$ the last two terms. Also, observe that since we are working under the assumption $J_i\to J^*$ almost surely   
as $i \to \infty$, it holds that 
\begin{align}
    \label{eq:convs}
(n-1)\det(B),\ (n-1)\tr(B)\to 0
\end{align}
almost surely as $n\to \infty$; we will need these facts later.
Now we use a well-known formula for the determinant of a sum of $2\times 2$ matrices, and the mean value theorem or representation for the remainder term in Taylor's theorem as in
\begin{align*}
    \sqrt{x_0+h}=\sqrt{x_0}+(2c)^{-1/2}h,  \text{ for some }c\text{ between }\text{$x_0$ and $x_0+h$}. 
\end{align*}
In this way
\begin{align*}
\det(A+B)^{1/2}&=\big(\det(A)+\det(B)+\det(A)\tr(A^{-1}B) \big)^{1/2}\\
&=\det(A)^{1/2}+(2 c)^{-1/2} \big(\det(B)+\det(A)\tr(A^{-1}B) \big).
\end{align*}
Using this and (\ref{eq:convs}), we get
\begin{align*}
 (n-1)&\bigg(\det\Big(J^*+\frac1n\sum_{i=1}^n\Delta J_i\Big)^{1/2}-\det\Big(J^*+\frac1{n-1}\sum_{i=1}^{n-1}\Delta J_i\Big)^{1/2}\bigg)\\
 &= (n-1)\bigg(\det(A)^{1/2}-\det(A+B)^{1/2}\big)\\
 &= -(n-1)(2 c)^{-1/2} \big(\det(B)+\det(A)\tr(A^{-1}B) \big)\to 0,
\end{align*}
almost surely as $n\to \infty$, since 
$$
\abs{\tr(A^{-1}B)}\le \norm{A^{-1}}_{\infty}\tr(B).
$$
Combining the above estimates, we deduce
\begin{align*}
    \sqrt{\det(J_1+\ldots+J_n)}- \sqrt{\det(J_1+\ldots+J_{n-1})}- \sqrt{\det(J_n)}\to 0
\end{align*}
almost surely when $n\to \infty$, as desired.

\section{Mathematical remarks on dynamic programming}
\label{sec:math}

In this section, we review part of the mathematical background for the optimal control problem in question, see for example, \citet{evans10,bardic97}. The literature is vast and it is impossible to give a thorough overview here. 

\subsection{Given final time, maximize $D$ problem}

\renewcommand{\t}{{\mathbf t}}

 Let $C_s>0$. The function $u:(0,\infty)\times [0,T]\to \R$ denotes the optimal amount $D$ we can accumulate during the remaining time if updating at the considered time. To be more precise, 
the controller chooses the update times 
\begin{align*}
 \t:=(t_0,\ldots,t_{n+1}),   
\end{align*}
 where the difference between the update times is larger than $C_s$, and otherwise the number $n$ may vary. Also, observe that $t_0$ and $t_{n+1}$ must correspond to the interval we are considering. For example, if we consider the full interval, then  $t_0=0$ and $t_{n+1}=T$, but this will be clear from the context.
The definition of $u$ now reads as
\begin{align}
    \label{eq:def-of-u}
   u(D,t)&=\begin{cases}
        \sup_{\t} \sum_{i=0}^n (t_{i+1}-t_i-C_s) h(D(t_i)) ,& t<T-C_s\\
   0,& t\in [T-C_s,T],
    \end{cases}
\end{align}
where
\begin{align}
\label{eq:dynamics}
        D'(s)&=\begin{cases}
    h(D(t_i)),&\text{ if }s\in [t_i+C_s,t_{i+1}),\\
    0,& \text{ otherwise}.
\end{cases}
\end{align}
Obviously, $u$ given by (\ref{eq:def-of-u}) exists and is unique.
\begin{lemma}[DPP]
    Let $u$ be as in (\ref{eq:def-of-u}). Then for $C_s>0$ and $t<T-C_s$ 
    \begin{align}
\label{eq:dpp}
 u(D,t)&=\sup_{t_{\text{next}}\in (t+C_s,T]}
\Big\{ u(D_{\text{next}},t_{\text{next}})+(D_{\text{next}}-D)\Big\},
\end{align}
where
\begin{align*}
D_{\text{next}}:&=D+h(D)(t_{\text{next}}-t-C_s).
\end{align*}
\end{lemma} 
The validity of the DPP can be readily verified by plugging in the definition (\ref{eq:def-of-u}). One could also study the DPP on its own right without invoking  (\ref{eq:def-of-u}) and showing existence and uniqueness using a similar idea as in the numerical computation: since $C_s>0$, iterating backwards from the end $T$, we can extend the values to each strip of width $C_s$.

\begin{remark}[Regularity, stability of update times]
\label{rem:regularity-stability}
Similarly as in  \citet[Section 10.3]{evans10}, we can also prove that $u$ is bounded and Lipschitz continuous in $(0,\infty)\times [0,T]$.  
From the definition of $h$ and \eqref{eq:def-of-u}, it is straightforward that
$$ u(D,t)\le h^{\ast}(T-C_s),$$
where $h^{\ast}=\sqrt{\det(J^{*})}=0.80940268$ is the maximum value of $h$.

The basic idea in the proof of Lipschitz continuity is to `copy' the update times from a nearby point of interest.
We will show
$$|u(D,t)-u(\tilde{D},\tilde{t})|\le C(|D-\tilde{D}|+|t-\tilde{t}|)$$
for some universal $C>0$ and any $(D,t),(\tilde{D},\tilde{t})\in (0,\infty)\times [0,T]$.
We first estimate  $$
|u(D,t)-u(\tilde{D},t)|
$$
and without loss of generality we may assume that $\tilde D\ge D$ and  $u(\tilde{D},t)\ge u(D,t)$.
To this end, for $t\in[0,T]$, we can choose update times $\mathbf{t}_1=(t_0,\dots,t_n)$ with $t=t_0$ which satisfies 
$$ u(\tilde{D},t)\le   \sum_{i=0}^n (t_{i+1}-t_i-C_s) h(\tilde{D}(t_i)) +\eps$$ for a fixed $\eps>0$.
Then we have
\begin{align*}
 & u(\tilde{D},t)  -u(D,t) \\ &\le
  \bigg( \sum_{i=0}^n (t_{i+1}-t_i-C_s) h(\tilde{D}(t_i)) +\eps\bigg)-\sum_{i=0}^n (t_{i+1}-t_i-C_s) h(D(t_i)) 
  \\ & = \sum_{i=0}^n (t_{i+1}-t_i-C_s) \big(h(\tilde{D}(t_i))-h(D(t_i))\big) +\eps.
\end{align*}
Since
\begin{align*}
 (\tilde{D}-D)'(s)  &=\begin{cases}
   h(\tilde{D}(t_i))-  h(D(t_i)),&\text{ if }s\in [t_i+C_s,t_{i+1}),\\
    0,& \text{ otherwise}
\end{cases}
\end{align*}
and $h$ is Lipschitz,
we observe that for any $s\in(t,T]$,
$$| (\tilde{D}-D)(s)| \le C|\tilde{D}-D|$$
for some $C>0$ independent of $\eps$ by Gronwall's inequality.
By employing this and Lipschitz regularity, we can see that
$$  u(\tilde{D},t)-u(D,t)  \le C|\tilde{D}-D|$$ for some $C>0$.

To get an estimate for $t$-direction, it is enough to show
$$u(D,t)-u(D,\tilde{t})\le C(\tilde{t}-t)$$ for $t<\tilde{t}$, since $u$ is decreasing in $t$. 
We use update times $\mathbf{t}_2=(s_0,\dots,s_k),\ s_0=t$ such that
$$ u(D,t)\le  \sum_{i=0}^k (s_{i+1}-s_i-C_s) h(D(s_i)) +\eps$$ for some $\eps>0$.
We set update times $\mathbf{\tilde{t}}_2=(\tilde{s}_0,\dots,\tilde{s}_k)$ so that updates always occur at the same values of $D$ as in the process starting at $t$ before hitting $T$,
that is, 
$$    \tilde{s}_i = \min \{
        \tilde{t}-t+s_i ,T \} $$
for each $i=0,\dots,k$.
Now we have
\begin{align*}
&u(D,t)-u(D,\tilde{t})\\ & \le 
\bigg( \sum_{i=0}^k (s_{i+1}-s_i-C_s) h(D(s_i)) +\eps \bigg)
-\sum_{i=0}^k (\tilde{s}_{i+1}-\tilde{s}_i-C_s) h(D(s_i))
\\ & \le h^{\ast} (\tilde{t}-t)+\eps
\end{align*}
and then we can complete the proof by letting $\eps \to 0$.

 This proof also has an implication that may be of independent interest: Since the copied strategy provides almost as good (in Lipschitz sense) outcome in the nearby points, small variations (numerical and other approximation errors) in the update times are not crucial. In other words, the update times are stable in this sense and this provides a kind of theoretical basis for the practical solving of the problem.   

It is also rather easy to see that 
\begin{align}\label{bound}
h(D)(T-t-C_{s})\le u_{C_{s}}(D,t) \le u_{0}(D,t+C_{s}),
\end{align}
where the subindex in $u_{C_{s}}$ and $u_0$ denote the solution to (\ref{eq:dpp}) with update cost $C_s$ or $0$. 
Intuitively, we can understand the upper bound in \eqref{bound} as the case of updating $D$ every time with $C_s=0$. Similarly, the lower bound can be regarded
as the case of no update for $D$.   
\end{remark}

Next, we observe that we cannot derive a partial differential equation from the DPP by passing to a limit with step size as usual, due to the lower bound given by the update cost $C_s$. Nonetheless, we can derive a partial differential equation if we consider a continuous time approximation with no update cost, and this can be solved explicitly.  
In this case, we can heuristically expect that updating at every time step is the optimal strategy for maximizing the $D$-value since there is no penalty for the updating  . 
Recall that  $u(D,t)$ denotes the amount of $D$ we can optimally accumulate during the rest of the experiment if starting at $(D,t)$ and updating immediately. When $C_s=0$, and the time step is 1, it holds that 
\begin{align}\label{wts}
u(D,t)
=  u(D+h(D),t+1)+h(D),
\end{align}
which is easy to show by induction.
This means that updates of $D$ occur after every measurement , that is, our heuristic is indeed true .
We can instead of \eqref{wts},  consider the following approximation where the time step is $\eps>0$  
\begin{align*}
u(D,t)=u(D+\eps h(D),t+\eps)+\eps h(D),
\end{align*}
and then pass to a continuous time limit. From this we obtain
\begin{align*}
\frac{u(D+\eps h(D),t+\eps)-u(D,t)}{\eps}= -h(D),    
\end{align*}
and if $u\in C^1$ passing to $\eps\to 0$, we get 
\begin{align}\label{mdeq}
h(D)u_{D}(D,t) + u_{t}(D,t) = -h(D) \qquad \textrm{in} \ (0,\infty)\times(0,T)
\end{align} 
with a boundary condition $u(D,T)= 0$, where now the time variables are continuous. This equation could be solved by the standard method of characteristics, see, for example, \citet{evans10}. 
The limit is also a solution to the DPP for all $\delta>0$ such that $0\le t, t+\delta\le T$  
\begin{align*}
&\begin{cases}
u(D,t)&=\int_t^{t+\delta} h(D(s)) \ud s+u(D(t+\delta),t+\delta),\\
u(D,T)&=0,\\
\end{cases}
\end{align*}
with the dynamics
\begin{align*}
&D'(s)=h(D(s)),\quad D(t)=D.    
\end{align*}
In this case, it might be instructive to think $h(D)$ as a density of accumulation of data so that $h(D(s))\ud s$ gives the accumulated data over a short time interval.

\subsection{Given desired $D_{\text{final}}$, minimize the time cost}

\renewcommand{\D}{{\mathbf D}}

Here we consider a continuous time version of (\ref{eq:dppt-discr}) i.e., $v$ now denotes the minimal time cost to reach the given $D_{\text{final}}$ if updating immediately. This time cost does not depend on the time spent so far but only $D$ accumulated so far (this is 'model $D$' obtained using $h(D)$ with $(a,b)=(1,0)$). Thus $v$ does not depend on time. We denote 
\begin{align*}
    \D=(D_0,\ldots,D_{n+1})
\end{align*}
where $D_0$ and  $D_{n+1}$ correspond to the interval we are considering.  Now
$v:(0,\infty)\to [0,\infty)$ can be defined as
\begin{align}
    \label{eq:def-of-v}
   v(D)&=\begin{cases}
        \inf_{\D} \sum_{i=0}^n \big\{C_s+(D_{i+1}-D_i) h^{-1}(D_i)\big\},&  D<D_{\text{final}}\\
   0,& t\in [D_{\text{final}},\infty).
    \end{cases}
\end{align}
Also observe that we can switch between $\t$ and $\D$: moving from $\t$ to $\D$, recall that $D(s)$ may be solved from (\ref{eq:dynamics}),
and thus we can solve $D_i$ corresponding to $t_i$. Knowing $D(t)$, we may also solve the inverse function $t(D)$ by selecting, for example the left continuous version of it at update points where it jumps. Using this function we may then move from $\D$ to $\t$. Again $v$ defined by (\ref{eq:def-of-v}) exists, is unique, and satisfies the DPP.
\begin{lemma}[DPP]
\label{lem:dppt}
    Let $v$ be as in (\ref{eq:def-of-v}). Then
\begin{align}
\label{eq:dppt}
v(D)&=\inf_{\Delta t\in (C_s,\infty)}
\Big\{ v\big(D+(\Delta t-C_s)h(D)\big)+\Delta t\Big\},\\
v(D)&=0\quad \text{ for }\quad D\ge D_{\text{final}}.\nonumber
\end{align}
\end{lemma} 
We see the validity of the DPP directly by plugging in (\ref{eq:def-of-v}).

\begin{remark}[Regularity, stability of update times]
\label{rem:regularity-stability-v}
The solution $v$ is bounded and Lipschitz continuous in $(0,\infty)$.
The idea is again to 'copy' the update points from a nearby point of interest. Similarly, the proof also implies that the update points are stable which is important for approximations.   

We also see that the following estimate holds 
$$v(D) \le \frac{D_{\text{final}}-D}{h(D)}+C_s$$
by considering the situation that an update is done only once.
\end{remark}

When $C_s=0$, and the time step is 1, we update $D$ every time to minimize the time cost so that the discrete version  (\ref{eq:dppt-discr}) becomes
\begin{align*}
 v(D)&= v(D+h(D))+1
\end{align*}
since $h$ is increasing. In order to pass to a continuous time, we can consider
\begin{align*}
 v(D)&= v(D+h(D)\eps)+\eps
\end{align*}
for  $\eps>0$. By passing to a limit $\eps\to 0$, we obtain the equation
\begin{align}\label{dfeq}
\begin{cases}
    h(D)v_{D}(D)&=-1 \quad \textrm{in}\ (0,D_{\text{final}})\times(0,\infty)\nonumber\\
v(D_{\text{final}})&=0.
\end{cases}
\end{align}
The solution also satisfies the DPP
\begin{align*}
&\begin{cases}
v(D)&=t+h+v(D(t+h)),\\
v(D_{\text{final}})&=0,\\
\end{cases}
\end{align*}
with the dynamics
\begin{align*}
&D'(s)=h(D(s)),\quad D(t)=D,\quad D<D_{\text{final}}.    
\end{align*}

\bigskip

\section*{Supplementary Materials}

 This supplement contains the R files for reproducing the simulations presented in the paper. A readme file is provided to connect the files with the corresponding simulations.
\par
\section*{Acknowledgements}

J. Han was supported by Basic Science Research Program through the National Research Foundation of Korea (NRF) funded by the Ministry of Education (NRF-2021R1A6A3A14045195).
\par

\section*{Declaration of Competing Interests}
The authors declare that there is no conflict of interest regarding the publication of this article.

\end{appendices}

\end{document}